\documentstyle[aps,epsfig]{revtex} 
\begin{document}
\title{  Jet  quenching  and  neutral  pion  production  in Au+Au
collisions at RHIC }

\author{\bf A. K. Chaudhuri\cite{byline}}
\address{ Variable Energy Cyclotron Centre\\
1-AF,Bidhan Nagar, Kolkata - 700 064\\}
\maketitle
\begin{abstract}

In the jet quenching model, we have analysed the PHENIX 
data  on  nuclear  modification  factor  of $\pi^0$, in  Au+Au  collisions at
$\sqrt{s}$=200  GeV, and extracted the initial gluon density of the 
medium produced.
In jet quenching, partons lose energy in the
medium, depending on the
medium density as well as on the in-medium path length. 
Systematic  analysis indicate  that  in  most
central (0-10\% centrality) collisions, medium density  is
very   large   $dN_g/dy   \sim$  2150.  Medium density decreases
exponentially as the collision centrality decreases and  in  very
peripheral  (70-92\%  centrality)  collisions,  $dN_g/dy \sim$ 3.
Initial energy density of the medium also decreases smoothly from $\varepsilon_0
\sim$ 20 $GeV/fm^3$ in most central collisions to  $\varepsilon_0
\sim$  3  $GeV/fm^3$  in  most  peripheral collisions. Very large
$dN_g/dy$ or $\varepsilon_0$ indicate very dense matter
formation in central Au+Au collisions.
\end{abstract}


\section{Introduction}

One  of the predictions of quantum chromodynamics is the possible
existence  of  a  deconfined  state  of  quarks  and  gluons.  In
relativistic  heavy  ion  collisions,  under  certain  conditions
(sufficiently  high  energy  density  and  temperature)  ordinary
hadronic  matter  (where  quarks  and  gluons  are  confined) can
undergo a phase transition to a deconfined matter, commonly known
as quark gluon plasma (QGP). Over the years,  nuclear  physicists
are  trying to produce and detect this new phase of matter, first
at CERN SPS and now at Relativistic Heavy Ion Collider (RHIC). At
RHIC several new results are obtained, most important among  them
is the high $p_T$ suppression in Au+Au collisions and its absence
in  d+Au  collisions.  All  the  four  experiments, STAR, PHENIX,
PHOBOS and BRAHMS  \cite{star1,phenix1,phobos1,brahms1}  observed
that   high  $p_T$  hadron  production  in  Au+Au  collisions  at
$\sqrt{s}=200$  GeV  (as  well  as  at  $\sqrt{s}=130$GeV),  {\em
scaled} by the average number of binary collisions, is suppressed
with respect to pp collisions. The suppression is more in central
collisions  than  in  peripheral  collisions. The observation has
created an excitement in nuclear physics community, as a possible
signature, if not of QGP, of creation of highly dense  matter  in
central Au+Au collisions.

One of the explanations offered for the high $p_T$ suppression at
RHIC  is the jet quenching \cite{gy90,wa92}. In the fragmentation
picture \cite{co82}, the single  parton  spectrum  is  convoluted
with  the probability for a parton $i$ to hadronize into a hadron
$h$, which carries a fraction $z < 1$  of  the  momentum  of  the
parent parton. Unlike in pp collisions, in AA collisions, partons
produced  in the initial hard scattering has to propagate through
a medium before  fragmentation.  While  propagating  through  the
medium,  the  partons  can  suffer  energy  loss through multiple
scattering. As a result, at the time of fragmentation to hadrons,
the partons will have less energy and consequently  will  produce
less  hadrons  than  would  have  otherwise. Several authors have
studied high $p_T$ production at RHIC energy in Au+Au  collisions
\cite{wa02,wa04,wa96,je03,mu03,vi02,wa03,vi04}.  The  LO  or  NLO
pQCD models with jet quenching can well explain  the  high  $p_T$
suppression,  implying  large  parton energy loss or high initial
gluon density. In \cite{wa02} gluon density, at an  initial  time
$\tau_0$= 0.2 fm was estimated to be 15-30 times higher than that
in a normal nuclear matter.

Apart  from  the jet quenching, alternative explanations are also
offered for the high $p_T$ suppression. Indeed,  leading  hadrons
from  the jet fragmentation can possibly have strong interactions
with the medium created and possibly be absorbed or its $p_T$  be
shifted to lower values leading to the high $p_T$ suppression. In
\cite{ga03}  high  $p_T$ suppression, due to final state hadronic
interactions, was considered. It was argued that  partons  cannot
materialize into hadrons in a deconfined phase. Fragmentation can
occur  only  outside  the  deconfined  phase or in vacuum. It was
argued that high $p_T$ (pre-)hadrons can well be realized  inside
the   late  stage  of  the  fireball.  Then  interaction  of  the
(pre-)hadrons  with  bulk  hadronic  matter  could  lead  to  the
observed  suppression. RHIC data on high $p_T$ suppression can be
partially  explained  in  the   model.   Final   state   hadronic
interaction  was also used by Capella et al\cite{ca04} to explain
the high $p_T$ suppression. In their model,  high  $p_T$  hadrons
interact  with the comovers and their $p_T$ is shifted to smaller
values.  Due  to  steepness  of the $p_T$ distribution the effect
could be large. Indeed, it was shown that the RHIC data are  well
explained   in   such  a  model  \cite{ca04}.  Recently  Xin-Nian
\cite{wa03} has argued strongly against models  using  the  final
state   hadronic   interaction   to   explain   RHIC  high  $p_T$
suppression. He argued that hadron formation time  could  not  be
small,  rather  large, in the range of 30-40 fm, much larger than
the  typical  medium  size  or  lifetime  of  the  dense  medium.
Moreover,   nearly   flat   $p_T$   dependence  of  the  observed
suppression at high $p_T$ empirically leads to  a  linear  energy
dependence   of  the  hadronic  energy  loss.  Since  the  hadron
formation time is proportional to hadron or jet energy, the total
energy loss due to hadron rescattering  or  absorption  decreases
with  energy.  He  argued that observed high $p_T$ suppression at
RHIC could only be due to partonic energy loss.

In  the  present  paper,  a systematic analysis of PHENIX data on
$\pi^0$ suppression  is presented. Data are analysed in the jet quenching
model.   In jet quenching, partons suffer  energy  loss  in
the medium produced in Au+Au collisions. 
  Partonic energy loss depend on the in-medium path
length  and on the initial gluon density \cite{gy90}. For a reliable estimate
of the gluon density, in-medium path length should be accurate.
In the present work,  in-medium  path  length,  in  each
centrality  ranges  of  collisions,  is  estimated in the Glauber
model. Initial gluon density in different centrality ranges of collisions are the
extracted from a  fit  to
the  PHENIX  $\pi^0$  data on the nuclear modification factor.  With  a  single parameter, the gluon
density,  large  $p_T  >  4 GeV$ part of the PHENIX data are well
explained.  Our  analysis  indicate  very  large  initial   gluon
density,  $dN_g/dy  \sim  2158$  in 0-10\% centrality collisions.
Corresponding energy density $\varepsilon_0  \sim  20  GeV/fm^3$.
Gluon  density decreases exponentially from central to peripheral
collisions  and   in   very   peripheral   (70-92\%   centrality)
collisions. it is very small, $dN_g/dy \sim 3$.

The  plan  of  the  paper  is  as  follows: In section 2, we have
described the pQCD model for hadron  production.  In  section  3,
results of our analysis of PHENIX data are presented. Summary and
conclusions are given in section 4.

\section{pQCD model for hadron production}

Details of high $p_T$ hadron production in pp collisions could be
found in \cite{eskola03}. In a pQCD model, the differential cross
section  for  production  of  a hadron h with transverse momentum
$q_T$ , at rapidity $y$, in pp collisions can be written as ,

\begin{eqnarray} \label{1}
\frac{d\sigma^{pp->hX}}{dq^2_Tdy} = &&K J_(m_T,y) \int \frac{dz}{z^2}
\int dy_2  \sum_{<ij>,<kl>} \frac{1}{1+\delta_{kl}} \frac{1}{1+\delta_{ij}}\\
\nonumber
&& \{ x_1 f_{i/A}(x_1,Q^2) x_2 f_{j/B}(x_2,Q^2)
\left[\frac{d{\hat \sigma}}{d{\hat t}}^{ij->kl}(\hat s, \hat t, \hat u) D_{k->h}(z,\mu^2)
+ \frac{d{\hat \sigma}}{d \hat t}^{ij->kl}(\hat s,\hat u,\hat t) D_{l->h}(z,\mu^2)\right]\\
\nonumber
&&+x_1 f_{j/A}(x_1,Q^2) x_2 f_{i/B}(x_2,Q^2)
\left[\frac{d{\hat \sigma}}{d \hat t}^{ij->kl}(\hat s,\hat u,\hat t) D_{k->h}(z,\mu^2)
\frac{d{\hat \sigma}}{d\hat t}^{ij->kl}(\hat s,\hat t,\hat u) D_{l->h}(z,\mu^2)\right]
\}
\nonumber
\end{eqnarray}

Details  of  the  equation can be found in \cite{eskola03}. Here,
the parameter K takes into account the higher order  corrections.
For  the partonic collisions $ij \rightarrow kl$, $x_1$ and $x_2$
are the fractional momentum of the colliding partons $i$ and $j$,
$x_{1,2}=\frac{p_T}{\sqrt{s}}(e^{\pm  y_1}+e^{\pm  y_2})$,  $y_1$
and $y_2$ being the rapidities of the two final state partons $k$
and  $l$.  For  the parton distribution, $f(x,Q^2)$, we have used
the CTEQ5L parton distribution, with the factorization scale $Q^2
\approx p_T^2$. In Eq.\ref{1},  $d{\hat  \sigma}^{ab->cd}/dt$  is
the  sub  process Cross-section. Only 8 sub processes contribute,
they   are  listed  in  \cite{sarcevic}.  $D_h(z,\mu^2)$  is  the
fragmentation function for the final state partons. We have  used
KPP  parameterization  \cite{KPP} with fragmentation scale $\mu^2
\approx  q_T^2$.  The  integration  region  for   $y_2$,   $-\ln(
\sqrt{s}/p_T  -e^{-y_f})  <y_2  <\ln( \sqrt{s}/p_T -e^{y_f})$, is
over the whole phase space, whereas that for z is

\begin{equation}   \frac{2m_T}{\sqrt{s}}   cosh  y  \leq  z  \leq
min[1,\frac{q_T}{p_0} J(m_T,y)] \end{equation}

\noindent with, $J(m_T,y)=(1-\frac{m^2}{m_T^2coshy})^{-1}$. $p_0$
is  the  cut  off  used to regulate infrared singularity. It is a
parameter and in the present work we have used $p_0$=1.0 GeV.

In Fig.1, we have shown the fit obtained to the PHENIX pp data on
$\pi^0$  production  with  K=$1.29\pm  0.02$.  LO pQCD model give
reasonably good description of the high $p_T$ $\pi^0$  production
in  pp collisions at RHIC energy. We note that better description
could  be   obtained   with   generalized   parton   distribution
\cite{ac03}, however, at the expense of an additional parameter.

Following  the  standard procedure, $p_T$ distribution of neutral
pions in Au+Au collisions could be obtained as,

\begin{equation}   \label{2}   \frac{d\sigma^{AA->hX}}{d^2q_Tdy}=
\int_{b_{min}}^{b_{max}} d^2b d^2s T_A({\bf b-s})T_B({\bf s})
\frac{d\sigma^{NN->hX}}{d^2q_Tdy}
\end{equation}

The impact parameter integration ranges ($b_{min}$ and $b_{max}$)
are chosen according to centrality of collisions. They are listed
in  table  1.  In  this  picture,  all the nuclear information is
contained in the thickness function, $T_A({\bf b})=\int \rho({\bf
b},z)  dz$.  We  have  used  Woods-Saxon  form  for  the  density
$\rho(r)$,

\begin{equation}             \rho(r)=\frac{\rho_0}{1+e^{(r-R)/a}}
\end{equation}

\noindent  For  Au,  $R=6.38  fm$  and  $a=0.535 fm$. The central
density $\rho_0$ is  obtained  from  the  normalizing  condition,
$\int \rho(r) d^3r =A$.

In AA collisions, vacuum parton distribution ($f(x,q^2)$) as well
as  the fragmentation function ($D_h(z,\mu^2)$) will be modified.
It is well known that parton distribution in bound nucleon differ
from that of a free nucleon (the EMC effect). We take care of  it
using  HIJING parameterization. However, its effect on high $p_T$
suppression  is  small.  Medium  modification  of   fragmentation
function  is  not very clear. Ideally, one should solve the DGLAP
evolution  equation  for  fragmentation  functions,  taking  into
account  the  medium  effects on the splitting function. Such an
analysis has not been performed. In the jet quenching  or  energy
loss model, high energy partons loses a fraction $\varepsilon$ of
its energy while passing through the medium and then fragments in
the vacuum with the normal vacuum fragmentation function with the
corresponding  shifted  momentum  fraction.  Any  modification of
virtuality dependence of the fragmentation function is neglected. The
medium modified fragmentation function is then written as,

\begin{equation}
D^{med}_{i\rightarrow h}(z,\mu^2) \rightarrow \int d\varepsilon p(\varepsilon)
\frac{1}{1-\varepsilon} D_{i\rightarrow h}(\frac{z}{1-\varepsilon},\mu^2)
\end{equation}

\noindent where $p(\varepsilon)$ is the probability that prior to
hadronisation,  a  parton  with  energy  $E$ loss $\varepsilon E$
energy in the medium. In the Poisson approximation of independent
gluon emission, probability  distribution  of  fractional  energy
loss  $p(\varepsilon)$  can  be  obtained iteratively from single
inclusive gluon radiation spectrum $dN/dx$. In the present paper,
we however choose a simpler Delta distribution, $p(\varepsilon) =
\delta(\varepsilon)$. Delta function  produces  a  much  stronger
effect due to rapid $z$ dependence of the fragmentation function.
With  delta  function  for  $p(\varepsilon)$,  implementation  of
energy loss is  simple.  For  $\Delta  E$  loss  of  energy,  the
fragmentation function in the elemental cross section, is changed
as,

\begin{equation} \label{11}
z D_h(z) = z^* D_h(z^*), \hspace{1.5cm} z^*=z/(1-\Delta E/E)
\end{equation}

As  the  fragmentation  functions  are  peaked at low z, partonic
energy loss will  necessarily  lead  to  less  number  of  hadron
production.

At  RHIC,  nuclear effects on the inclusive spectra, in different
centrality ranges of collisions, were measured in  terms  of  the
nuclear modification factor ($R_{AA}$). It is defined as,

\begin{equation} \label{5}
R_{AA}=\frac{d^2N/dp_Tdy (Au+Au)}{T_{AA}d^2\sigma/dp_Tdy  (pp)}
\end{equation}

\noindent  where  $T_{AA}=<N_{bin}>/\sigma_{inel}$ from a Glauber
calculation accounts for  the  nuclear  collision  geometry.  The
nuclear  modification  factor  will  be  unity, had there been no
nuclear effects on high $p_T$ production. In Fig.2,  PHENIX  data
on  nuclear  modification  factor for neutral pions, in different
centrality ranges of  collisions  are  shown.  PHENIX  data  show
considerable   nuclear   effects   ($R_{AA}   <  1$)  in  central
collisions.

\section{Partonic energy loss}

In  the jet-quenching picture, hadronic spectra are determined by
the energy loss of partons. Partonic energy loss in a medium is a
complex phenomena. Recently,  much  progress  has  been  made  in
understanding            partonic           energy           loss
\cite{gy94,ba97,ba00,gy00,gy00a,we00,gu00}. Partons  lose  energy
mainly  due  to  gluon  radiation. QCD radiative energy loss in a
finite size QGP was solved analytically in  \cite{ba97}.  In  the
leading log approximation, BDMS \cite{ba97} prediction is ,

\begin{equation} \label{bdms}
\Delta E_{BDMS} = \frac{C_F \alpha_s}{4} \frac{L^2\mu^2_D}{\lambda_g} \tilde{v}
\end{equation}

\noindent  where  $L$ is the length of the plasma, $\mu_D$ is the
Debye screening mass, $\lambda_g$ is the mean free  path  of  the
gluons  in  QGP,  $C_F$  is  the  color  Casimir  for the partons
($C_F$=3  for  gluons  and  4/3  for  quarks)  and   the   factor
$\tilde{v}$  grows  smoothly  with  $L$,  at  $L  >>  \lambda_g$,
$\tilde{v} \approx \log(L/\lambda_g)$.

BDMS  energy  loss  is  independent  of  parton energy. Recently,
Gyulassy, Levai  and  Vitev  (GLV)  \cite{gy00},  calculated  the
partonic  energy  loss  in the light-cone path integral approach.
GLV form of the partonic energy loss depends  logarithmically  on
the  parton  energy. Discrepancy between the energy dependence of
the radiative energy losses in two approaches is understood to be
due to leading log approximation in BDMS approach \cite{za00}.

Gluon  radiation  from  a  static  source  has  a  characteristic
quadratic  dependence  on  the  in-medium   path   length   (L)
Eq.\ref{bdms}.  The  path  dependence is changed if the medium is
expanding. For example, in an one dimensionally expanding medium,
the quadratic  dependence  is  diluted  to  a  linear  dependence
\cite{gy00}.  Under  certain  approximations, analytic expression
for  partonic  energy  loss,  in  a  medium  undergoing   Bjorken
expansion , was obtained by Gylusaay et al \cite{gy00}.

\begin{equation} \label{vitev}
\Delta E = \frac{9C_F\pi\alpha_s^3}{4} \frac{1}{A_T} \frac{dN_g}{dy}
L \ln(\frac{2E}{\mu^2_DL}+\frac{3}{\pi})
\end{equation}

\noindent where as before $L$ is the in-medium path length and $A_T$
is  the  transverse  area.  It  can be seen that transverse gluon
density $\frac{1}{A_T}\frac{dN_g}{dy}$ as well as the medium path
length controls the partonic energy loss.

In  addition  to  gluonic  radiation,  partons suffer collisional
energy loss. There are marked differences between  radiative  and
collisional  energy  loss.  Detailed  calculation  of collisional
energy loss of partons has not been made.  $\Delta  E_{coll}$  is
independent  of  in-medium path length and depend logarithmically
on energy. It is determined  mainly  by  the  medium  temperature
\cite{bj82},

\begin{equation}
\frac{dE_{coll}}{dz} \propto \alpha_s^2 T^2 ln E/T
\end{equation}

In general radiative energy loss dominate over collisional energy
loss.  However, for high $p_T>10 GeV$ partons, collisional energy
loss can  be  substantial  (30-40\%)  \cite{za04}.  Since  PHENIX
$\pi^0$  data are limited to $p_T <10 GeV$, we have neglected the
collisional energy loss.

\section{Results}

PHENIX   collaboration   measured   nuclear  modification  factor
($R_{AA}$) for $\pi^0$, in several centrality ranges of collisions.  
In  table
1,  we  have listed the centrality ranges of collisions analysed in the
present paper
  along with the  corresponding  impact  parameter  ranges
($b_{min}$ and $b_{max}$). For completeness purpose, we have also
listed  the Glauber model calculation of average number of binary
collisions and number of participants in those  centrality  ranges
of  collisions.  In  the  jet  quenching  model, particle spectra
depend on the energy loss suffered by the  partons.  In  the  GLV
form,  energy  loss  depend  on (see Eq.\ref{vitev}), (i) the initial transverse gluon
density      (ii)   the    Debye
screening  mass   and (iii) the in-medium path length. We
have assumed that in a centrality  range  of  collisions,  partons
travel  an  average  distance  $L$  and suffer energy loss in a
medium     with     average  initial    transverse     gluon density
$\rho^{trans}=\frac{1}{A_T}                 \frac{dN_g}{dy}$.
Since energy loss depend on the product $L \rho^{trans}$,
accurate determination of 
the density of the medium requires accurate in-medium  path length.
We have calculated
the average in-medium path length in a Glauber model (see appendix).  
In-medium
path  length $L$ in different centrality ranges of collisions 
are
listed in table 1. In  0-10\%  central  collisions,  the  average
medium  path  length  is  5.17 fm. It is nearly a factor of 1.5 times smaller
than the length $4/3R_A$, used by Vitev et al \cite{vi02} in their calculation.
In-medium path length
reduces by a factor of 5 in
very peripheral collisions. For the screening mass
$\mu_D$ we use a fixed value $\mu_D$=0.8  GeV.  This specific value was obtained from a fit to
PHENIX 0-10\%  centrality  $\pi^0$  data,  with  both  the  gluon
density and $\mu_D$ as free parameters.

With $\rho^{trans}$ as the only free parameter, nuclear modification
factors for $\pi^0$ are fitted, using the MINUIT minimization programme.
pQCD is applicable only for high $p_T$. Consequently only a part
of the data, $p_T > 4 GeV$ 
are considered for the fitting. 
In Fig.2, PHENIX data along with the fit obtained are shown.
Within the
error bars, high $p_T$ part of the PHENIX  data  on  the  nuclear
modification  factor  are  well  described.  Energy  loss  or jet
quenching can explain the dramatic variation of nuclear  effects,
in  $\pi^0$  production,  in  terms  of  a  single parameter, the
average            transverse            gluon            density
$\rho^{trans}$.   
However, we note 
the
discrepancy in the $p_T$ dependence of the experimental  $R_{AA}$
and  the  pQCD  predictions. At large
$p_T$, within the error bars, experimental $R_{AA}$  donot  show  any $p_T$ dependence. The pQCD model
predictions, on  the  other  hand  show     $p_T$  dependence,
suppression decreasing with $p_T$.
  In the jet quenching models, relative energy
loss ($\Delta E/E$) is the relevant quantity. In GLV form of partonic
energy loss, $\Delta E/E \propto ln E/E$. As a
results, at large $p_T$, partons lose less energy and suppression
is   reduced.    PHENIX  data  have  large  error  bars,
reflective of poor statistics at RHIC.  Good  statistic  data  in
extended  $p_T$  range can resolve the issue, whether or not high
$p_T$ suppression is $p_T$ independent and test the GLV  form  of
partonic energy loss.

Initial transverse   gluon   density   ($\rho^{trans}$)   in   different
centrality ranges of collisions, as extracted from the PHENIX data, are  listed
in  table  1.  
In 0-10\% centrality Au+Au collisions,
gluon  density is very large, $\rho^{trans}_{0-10\%}>=21.08\pm
1.00$.  Density  decreases  as  the  centrality of the collisions
decreases.       In       very       peripheral       collisions,
$\rho^{trans}_{70-92\%}=4.5\pm1.22$,  nearly a factor of 5 less
than density in very central collisions.
In Fig.3, we have shown the extracted initial transverse gluon density,
as a function of the number of participants. The linear relation,

\begin{equation}
\frac{1}{A_T} \frac{dN_g}{dy} = 3.74 + 0.11 \frac{N_{part}}{2},
\end{equation}

\noindent  describe  the  extracted  initial transverse  gluon density in
different  centrality  ranges  of  collisions.  

Initial  rapidity  density  of  gluons  ($dN_g/dy$)  can  also be
estimated.  The  average  transverse  area   ($A_T$),   in   each
centrality  ranges  of  collisions,  is  easily calculated. At an
impact parameter $b$, the transverse area is elliptical and  $A_T
=\pi  a  b$ with $a=\sqrt{R^2-b^2/4}$ and $b=R-b/2$. We calculate
the average transverse area taking the simple average of $A_T$ at
two limits of the impact parameter ($b_{min}$ and $b_{max}$)  ranges.
In   table  1,  average  transverse  area  $A_T$  in  different
centrality ranges of collisions  are  listed.  $A_T$  decreases
very rapidly as the collision centrality decreases. While in very
central   collisions,  $A_T$  $  \sim  100  fm^{-2}$,  in  very
peripheral collisions, $A_T$ $< 1 fm^{-2}$. This  is  reflected
in  gluon rapidity density $\frac{dN_g}{dy}$ (see table 1). 
Gluon
density decreases exponentially  with  centrality.  In   0-10\%
centrality   collisions,  initial gluon  density,  $\frac{dN_g}{dy}  \sim
2158$,  and in very peripheral collisions $dN_g/dy \sim$ 3. Gluon
density in 0-10\% centrality collisions, is nearly  factor  of  2
larger  than the density obtained by Vitev et al \cite{gy00a}. In
0-10\%  central  collisions,  they   obtained,   $dN_g/dy   \sim$
1000-1200.  The  difference  is  mainly  due to difference in the
estimate of average  medium  path  length.  Average  medium  path
length ($<L>=4/3 R_A$) used in \cite{vi02} is nearly a factor of
1.5 larger than the Glauber model estimate used presently.

In  Fig.4,  we  have  shown  the  initial rapidity density of gluons as a
function of the average path length $<L>$ in  the  medium.  Gluon
density increases exponentially with the medium path length ($L)$
and are well described by the empirical relation,

\begin{equation}
\frac{dN_g}{dy} = A exp(\mu L)
\end{equation}

\noindent   with   $A=0.2$  and  $\mu=1.8  fm^{-1}$.  Exponential
dependence of initial gluon density on medium path length is very
interesting. It  is  indicative  of  exponential  loss  of  gluon
intensity in a medium, $I=I_0 exp(-\mu L)$.

We  have  also  estimated  the  initial  energy  density  in each
centrality ranges of collisions. If $<p>$ is the  mean  transverse
momentum of gluons, energy density is,

\begin{equation}
\varepsilon_0 \approx  \frac{<p>^2}{A_T} \frac{dN_g}{dy}
\end{equation}

At  RHIC,  mean  transverse momentum is $<p_T> \sim 0.4-0.5$ GeV.
Assuming   duality   between   particles   and    gluons,    $<p>
=<p_T>_{particle}$,   the initial  transverse  gluon  density  is  easily
converted into initial energy density. They  are  shown  in  Fig.5  as  a
function  of  participant  number.  In  most  central collisions,
energy density as high as 20 $GeV/fm^3$.  The  value  agree reasonably
 well with  X.N.  Wang's  estimate  that in central Au+Au collisions,
compared to cold nuclear matter ($\varepsilon_{cold} \sim $0.16 $GeV/fm^3$), 
100 times larger energy density is produced.
As
the collision centrality  decreases, energy density decreases linearly and
in very peripheral collisions $\varepsilon_0 \approx 4. GeV/fm^3$.
The empirical relation,

\begin{equation}
\varepsilon(N_{part})=3.84 + 0.11 \frac{N_{part}}{2} (GeV/fm^3),
\end{equation}

\noindent  well  describe  the  energy  density  as a function of
number of participants. We note that even in peripheral collisions,
the deduced energy density is very large compared to cold nuclear
matter ($\varepsilon_{cold} \sim $0.16 $GeV/fm^3$). Indeed, it is
larger than  the critical
energy density for QGP phase transition is $\sim$ 1 $GeV/fm^3$.
 
It is interesting to see the model prediction at LHC energy.
At  LHC, centre of mass energy is very large ($\sqrt{s}$=1.148 TeV).
High $p_T$ particles will
be more abundant and $R_{AA}$ could be measured  more  accurately
in  extended $p_T$ range.   In Fig.6,
we have shown the
predicted nuclear modification factor ($R_{AA}$) in
0-10\% centrality Au+Au collisions at
LHC energy,  for  an   initial transverse  gluon   density,
$\frac{1}{A_T}\frac{dN_g}{dy}$=20. 
Between  $p_T$=5  to  20  GeV,  suppression  is
reduced by more than a factor of two. 
For comparison, $R_{AA}$ at
RHIC  energy  is  also shown (the red line). Interestingly, for a
fixed gluon density, suppression is more at RHIC than at LHC.
We also note that whether or not GLV form of partonic energy loss give
correct $p_T$ dependence can be verified at LHC if $R_{AA}$ is measured
accurately in the $p_T$ range, 5-20 GeV.

In Fig.7, we have shown the predicted nuclear modification factor
($R_{AA}$),  at  a  fixed  $p_T$=5  GeV,  as  a  function  of the
initial transverse gluon  density, in 0-10\% centrality collisions. 
Both at  RHIC  and  LHC  energy,  as expected, $R_{AA}$
decreases with increasing gluon density. However, here again,  we
find  that suppression is less at LHC energy than at RHIC energy.
At  LHC,  30\%  larger  gluon  density  is  required  to   obtain
suppression similar to RHIC energy ($R_{AA} \sim 0.2$).

\section{Summary and Conclusions}

To  summarise,  we  have  presented  a systematic analysis of the
PHENIX data, on the nuclear modification factor  for  $\pi^0$'s,  in
Au+Au  collisions, at $\sqrt{s}$=200 GeV. Data are analysed in the
Jet-quenching model, with the GLV form for  the  partonic  energy
loss in an one dimensionally expanding medium.  GLV form of
Partonic  energy 
loss depend on the medium path length and
initial (transverse) gluon density. Medium path lengths
in different centrality
ranges of collisions are calculated in a Glauber model and 
initial
(transverse) gluon densities are extracted from a  fit  to  the  data.
From central to peripheral collisions, initial gluon density decreases exponentially.
In 0-10\% centrality collisions, present analysis give,  $dN_g/dy
\sim$ 2158, nearly a factor of 2 larger than the value obtained by
Vitev  el  al  \cite{gy00a}.  The  difference  is  mainly  due  to
different estimate of in-medium path length. We also find  that  the
extracted   gluon  densities  follow  the   empirical  relation
$dN_g/dy =0.2  exp(1.8  L)$, $L$ being the in-medium path length.
The  result  is  interesting.  Like
photons,  gluon  intensity  is  exponentially  attenuated  in the
medium, with attenuation coefficient $\mu=1.8 fm^{-1}$.  Assuming
gluon-hadron  duality,  we  have  also  estimated  initial energy
density  ($\varepsilon_0$).  In  0-10\%  centrality   collisions,
$\varepsilon_0$  is  large, $\sim$ 20 $GeV/fm^3$. It decreases to
$\varepsilon_0  \sim$  3$GeV/fm^3$   in   very   for   peripheral
collisions. Even in peripheral collisions, energy density is much larger
than in cold nuclear matter ($0.17 GeV/fm^3$).
We  have  also  given  prediction  for high $p_T$ suppression at LHC
energy. For the same gluon density,  high $p_T$ $\pi^0$'s are less 
suppressed at LHC energy than at RHIC energy.

\eject
\appendix
\section{Path length  in Glauber model}

The  average  length  traversed  by  a  parton  at certain impact
parameter can  be  obtained  in  the  Glauber  model.  At  impact
parameter  ${\bf  b}$,  the  positions  $({\bf  s},z)$ and $({\bf
b-s},z\prime)$ specifies the initial hard scattering of  partons.
The  length  traversed  by  the  partons,  after the initial hard
scattering can be calculated as,

\begin{equation}
L({\bf b,s},z,z\prime) = \frac{1}{\rho_0}( \int_z^\infty dz_A \rho({\bf s},z_A) +
\int_{z\prime}^\infty
dz_B  \rho({\bf b-s},z_B))
\end{equation}

\noindent where $\rho_0$ is the central density. Above expression
should  be averaged over all positions of initial hard scattering
with a weight of the nuclear densities to obtain the path length at
impact parameter {\bf b}.

\begin{equation}
L({\bf b})=\frac{\int d^2s \int dz \rho_A(s,z) \int dz\prime \rho_B(b-s,z\prime) L(b,s,z,z\prime)}{
\int d^2s \int dz \rho_A(s,z) \int dz\prime \rho_B(b-s,z\prime)}
\end{equation} 

\begin{table}
\caption{For each centrality   ranges  of  collisions we list the
impact parameter ranges ($b_{min}$ and $b_{max}$,
Glauber  model  calculation  of the average number of binary
collisions ($<N>_{coll}$), and medium path length  ($<L>$).
Average transverse area ($A_T$) calculated in a hard sphere model
are also listed. Transverse gluon density ($\frac{1}{A_T}\frac{dN_g}{dy}$ )
and gluon density ($dN_g/dy$) in each centrality ranges of collisions
are also shown.
}

  \begin{tabular} {cccccccc}\hline
  centrality & $b_{min}(fm)$ & $b_{max}(fm)$&$<N>_{coll}(<N>_{part})$&$<L>$ (fm)&$A_T (fm^{-2})$
 & $\frac{1}{AT}\frac{dN_g}{dy}(fm^{-2})$ &$\frac{dN_g}{dy}$\\ \hline
\hline
0-10\%    &0.0 &4.7&955.4(325.2)& 5.174&102.2 &$21.12\pm1.00$&$2158.3\pm102.1$ \\
10-20\%  &4.7 & 6.7&602.6(234.6) & 4.843 &64.0&$16.94\pm0.89$&$1083.9\pm57.1$\\
20-30\%  & 6.7 & 8.2 &373.8(166.6) & 4.396&43.9 &$13.13\pm0.81$&$576.2\pm36.6$\\
30-40\%  & 8.2 & 9.5&219.8(114.2) & 4.014&28.7 &$10.38\pm0.76$&$298.2\pm21.8$\\
40-50\%  & 9.5 & 10.6&120.3(74.4) &3.393 &17.3&$7.12\pm0.73$&$123.3\pm12.7$\\
50-60\%  & 10.6 & 11.6 &61.0(45.5)& 3.032 &8.7 &$6.26\pm0.80$&$54.8\pm7.0$\\
60-70\%  & 11.6 & 12.5 &28.5(25.7)& 2.276 &2.9 &$4.17\pm0.86$&$12.0\pm2.5$\\
70-80\%  & 12.5 & 13.4 &12.4(13.4) & 1.450 &$<.65$ &$5.07\pm1.25$&$<3.3$\\
70-92\%  &12.5 & 14.5 & 8.3(9.5)  & 1.164 &$<.65$ &$4.58\pm1.22$&$<3.0$\\
  \end{tabular}
\end{table}

\eject

\begin{figure}[h]
\centerline{\psfig{figure=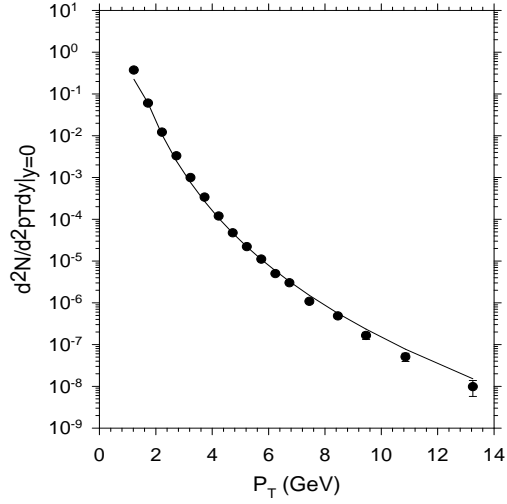,height=10cm,width=7.5cm}}
\vspace{-2cm}
\caption{The  transverse  momentum  spectra  for  the  neutral
pions, in PHENIX pp  collisions.  The  solid  line  is  the  pQCD model
calculation.} \end{figure}

\begin{figure}[h]
\centerline{\psfig{figure=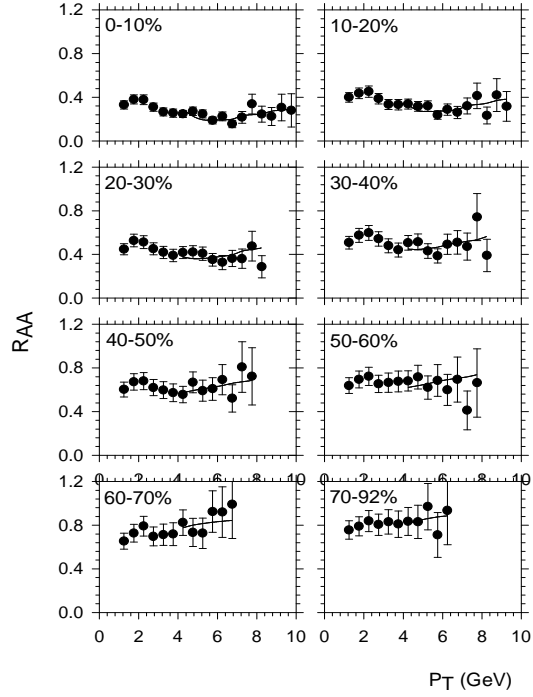,height=10cm,width=7.5cm}}
 \caption{The  PHENIX data on nuclear modification factor for the
neutral pions in different centrality of  collisions.  The  solid
lines are pQCD model fit to the data with jet quenching. Only free parameter
is the initial gluon density.}
\end{figure}

\begin{figure}[h]
\centerline{\psfig{figure=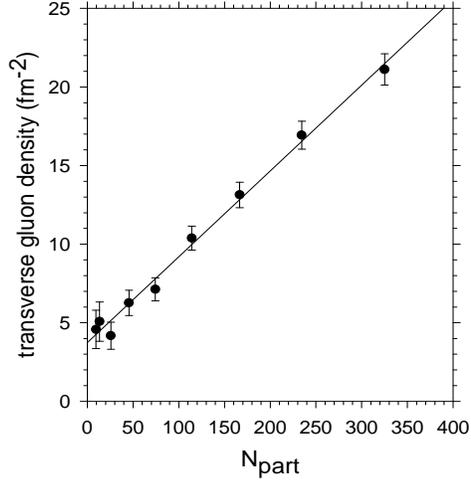,height=10cm,width=7.5cm}}
\vspace{-2.5cm}
 \caption{Initial transverse gluon density ($\frac{1}{A_T} \frac{dN_g}{dy}$) in different centrality
ranges of collisions as a function of number of participants.
Solid line is the linear relation, $\frac{1}{A_T} \frac{dN_g}{dy}=3.74 + 1.1\frac{N_{part}}{2}$.
 }
\end{figure}
\begin{figure}[h]
\centerline{\psfig{figure=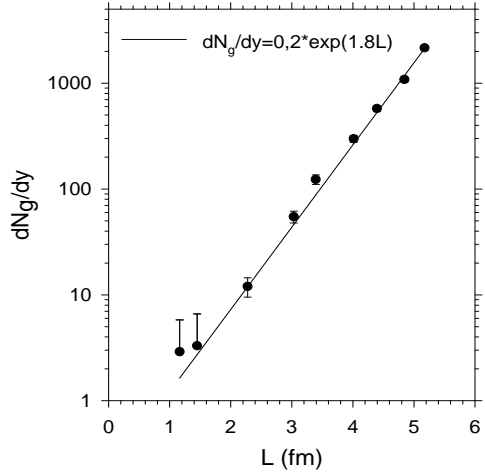,height=10cm,width=7.5cm}}
\vspace{-2.5cm}
 \caption{Initial rapidity density ($dN_g/dy$) of gluons, in different centrality
of collisions are shown as a function of medium path length (L).
The solid
line is a   fit,
$dN_g/dy=0.2 exp(1.8L)$.}
\end{figure}

\begin{figure}[h]
\centerline{\psfig{figure=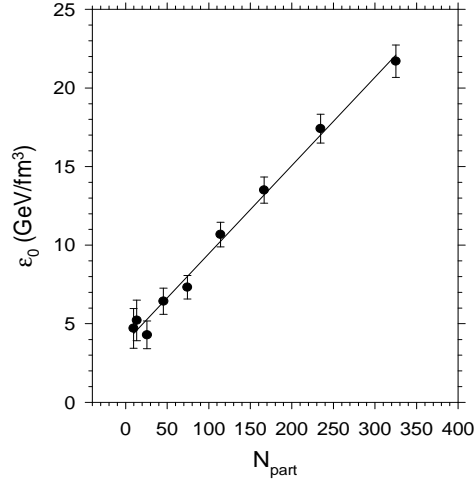,height=10cm,width=7.5cm}}
\vspace{-2.5cm}
 \caption{Initial energy density ($\varepsilon_0$ in different centrality
of collisions are shown as a function of participant numbers. The solid
line is a linear fit to the energy density,
$\varepsilon_0=3.84+1.1 \frac{N_{part}}{2}$.}
\end{figure}

\begin{figure}[h]
\centerline{\psfig{figure=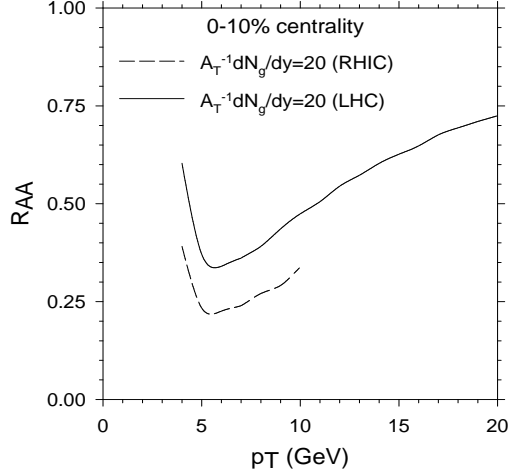,height=10cm,width=7.5cm}}
\vspace{-2.5cm}
\caption{Predicted nuclear modification factor in 0-10\% centrality Au+Au collisions
at LHC energy, shown as a function of transverse momentum. Initial transverse gluon density is $\frac{1}{A_T}\frac{dN_g}{dy}$=20. For comparison, the same at RHIC energy
is also shown (the dashed line).}
\end{figure}

\begin{figure}[h]
\centerline{\psfig{figure=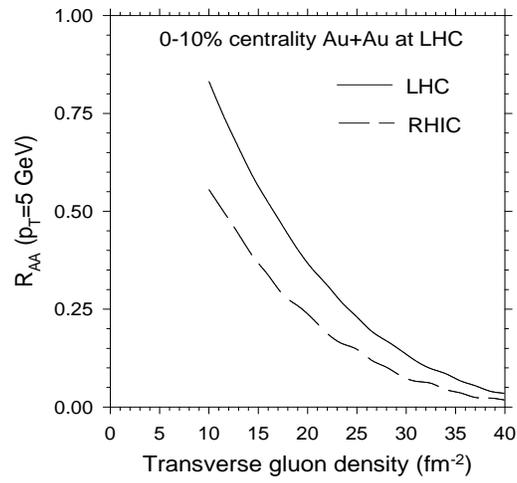,height=10cm,width=7.5cm}}
\vspace{-2.5cm}
 \caption{Predicted nuclear modification factor at $p_T$=5 GeV, in
0-10\% centrality Au+Au collisions. The solid line and the dashed lines
are for LHC and RHIC energy respectively.
}
\end{figure}
\end{document}